# Routing in FRET-based Nanonetworks

Pawel Kulakowski, Kamil Solarczyk, Krzysztof Wojcik

*Abstract*—Nanocommunications, understood as communications between nanoscale devices, is commonly regarded as a technology essential for cooperation of large groups of nanomachines and thus crucial for development of the whole area of nanotechnology. While solutions for point-to-point nanocommunications have been already proposed, larger networks cannot function properly without routing. In this article we focus on the nanocommunications via Förster Resonance Energy Transfer (FRET), which was found to be a technique with a very high signal propagation speed, and discuss how to route signals through nanonetworks. We introduce five new routing mechanisms, based on biological properties of specific molecules. We experimentally validate one of these mechanisms. Finally, we analyze open issues showing the technical challenges for signal transmission and routing in FRET-based nanocommunications.

## INTRODUCTION

WITH enormous growth and progress in the whole area of nanotechnology, the demand for communication between nanomachines has arisen naturally. Nanomachines, called also nanodevices or nanonodes, may be both of biological and artificial origin. Biological ones, e.g. proteins or whole cells, are building blocks of living organisms. Currently, scientists are working on artificial nanomachines, constructing structures like molecular switches, ratchets, and motors based on their biological counterparts [1]. The application field for nanomachines is extremely wide, extending from environment monitoring, industrial manufacturing, and building labs-on-a-chip to an enormous number of applications in medicine, like drug delivery, diagnostics, tissues regeneration, and surgery operations. The limited size of nanomachines, however, restrict their functions and capabilities, thus the ability to perform more complex actions rely on cooperation in larger groups, i.e.

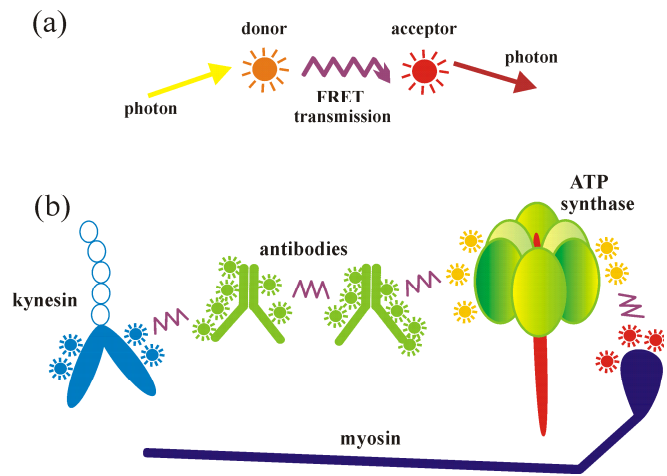

Figure 1. a) the FRET process after excitation of the donor molecule by an external photon. The excitation energy is passed non-radiatively to the acceptor molecule and then can be released as another photon.
b) examples of proteins that may work as nanomachines: antibodies, kynesin, ATP synthase and myosin molecules. Each of them has some fluorophores (marked as small circles with short rays) attached to it. The fluorophores serve as nanoantennas transmitting and receiving signals via FRET.

nanonetworks. In this sense, nanocommunications will play a similar role for nanotechnology, as telecommunications is currently playing for electronics: it will enable the nanomachines to work together. Nanodevices will communicate each other in order to: a) exchange and forward gathered information, b) coordinate their joint actions, c) interface with other biological and artificial systems. This is why the efficient communication between the nanodevices is a crucial challenge to be met in order to develop future nanonetworks and the whole area of nanotechnology.

The dimensions of nanoscale devices are comparable with single molecules, so the electromagnetic communication based on miniaturized transceivers and antennas is hard to be directly applied. Instead, numerous biologically inspired communication mechanisms have been proposed. Calcium signaling, molecular and catalytic nanomotors, pheromones propagation, and information transfer using bacteria as carriers have already been studied, but these mechanisms are slow and characterized by large propagation delays: the encoded data travels with a speed of several dozens of micrometers per second at maximum [2]. Comparing with these techniques, a mechanism based on the phenomenon of Förster Resonance Energy Transfer (FRET) is more promising, especially having in mind its low propagation delay [3–6]. FRET is a process in which a molecule, known as a donor, is able to non-radiatively (i.e. without releasing a photon) transfer its energy to another molecule, called an acceptor. For this energy transfer to occur,

Pawel Kulakowski is with Department of Telecommunications, AGH University of Science and Technology, Poland.
Kamil Solarczyk is with the Department of Cell Biophysics, Faculty of Biochemistry, Biophysics and Biotechnology, Jagiellonian University, Poland.
Krzysztof Wojcik is with II Chair of Internal Medicine, Faculty of Medicine, Jagiellonian University Medical College, Poland.

| Nanorouter type | Routing mechanism | Number of links outgoing from the nanorouter | What switches the nanorouter | Switching time |
|---|---|---|---|---|
| Proteins with multiple different fluorescent dyes | not needed | 2–4 | - | - |
| Photoswitchable fluorophores | change of absorption and emission spectra | 2 | light | 1–60 seconds |
| Quenchers | fluorescence quenching | 2–6 | temperature, pH | 1–10 seconds |
| Proteins with changeable shape | protein conformation change | 2 | adding a ligand | 10–100 milliseconds |
| ATP synthases | ATP synthase rotation | 2–10 | voltage, ATP | 2–50 milliseconds |

the donor must be in an excited state, the acceptor has to be located in close proximity to the donor and finally, the donor and acceptor molecules must be spectrally matched, i.e. the donor emission and acceptor absorption spectra (the frequency ranges of emitted/absorbed EM spectrum) should overlap (see Fig. 1a). The delay of the energy transfer is usually no more than 20 nanoseconds. The donor can be excited in various ways, e.g. by photon absorption, another FRET process or a chemical reaction (bioluminescence). The latter process, called Bioluminescence Resonance Energy Transfer (BRET) may be suitable for nanocommunications, as it does not require any external energy source, e.g. light, for donor excitation.

The FRET and BRET processes are very distance dependent: their efficiency decreases with the sixth power of the donor-acceptor separation. For each two types of molecules, the, so called, Förster distance can be experimentally measured; it is a separation where the respective FRET efficiency is equal to 50 percent. Förster distances for typical donor-acceptor pairs range from 3 to 9 nanometers. This effectively places a limit on FRET transmission distances to about 10 nanometers. Recent works [7–8] have shown that the FRET efficiency and transmission range can be additionally increased when using multiple donors at the transmitter side and multiple acceptors at the receiver side of the nanocommunication channel. Such a technique is called MIMO-FRET, following the idea of multiantenna systems well known in wireless communication. The most common molecules that can be used as FRET donors and acceptors are known as fluorophores. Their absorption and emission spectra lay in the visible light range, i.e. from 380 to 760 nm. It should be noted, however, that FRET as a phenomenon is not limited to this range of electromagnetic spectrum, but may occur also for shorter and longer wavelengths. When thinking about interactions of future nanodevices, one may envisage fluorophores performing functions of nanoantennas attached to some larger nanostructures, e.g. proteins like antibodies, myosins, kynesins, dyneins or ATP (adenosine triphosphate) synthases, working as nanomachines (see Fig. 1b).

The content of the article remainder is as follows. The next section introduces five new routing mechanisms available for FRET-based nanonetworks. Then, in order to prove its feasibility, the first of these mechanisms is validated experimentally and the results of the laboratory experiment are given. Later, the most important open issues in nanorouting are discussed and, finally, the article is concluded.

## ROUTING MECHANISMS

While the point-to-point communication via FRET, with its parameters like bit error rate, channel capacity, throughput, and signal propagation delay, was analyzed and measured in previous papers [5–8], the main motivation for this article is to go a step further and propose possible routing mechanisms in FRET-based nanonetworks. When thinking about fully operational nanonetworks, there must be not only point-to-point links, but also multi-hop connections. Nanonodes should be able to forward signals; moreover, the nodes should be able to take routing decisions, i.e. decide *where* to forward the signals. Assuming a nanonetwork communicates among its nodes via FRET, we propose five new techniques of signal routing. They are: (a) proteins with multiple different fluorescent dyes, (b) photoswitchable fluorophores, (c) quenchers (d) proteins with changeable shape, and (e) ATP synthases; they are described in the five following subsections and compared parametrically in Table 1. Until now, to our best knowledge there is still no research on these techniques in the area of nanocommunications. There are, however, numerous papers already published in life sciences where it was shown that these five mechanisms could provide suitable means for nanorouting. In the next section, we also report our laboratory experiments proving the efficiency of the first of the proposed routing techniques.

### Proteins with a few different fluorescent dyes

Attaching fluorescent dyes (fluorophores) to some larger molecules, e.g. proteins, is common in biology experiments in order to localize these proteins in their environment. For the purpose of nanocommunications, the fluorophores can serve as nanoantennas which are able to communicate with each other via FRET. Each fluorophore is characterized by its emission spectrum. The FRET signals emitted by this fluorophore can be received only by neighboring fluorophores with absorption spectra matching this emission spectrum.

Many types of fluorescent dyes can be attached to a single

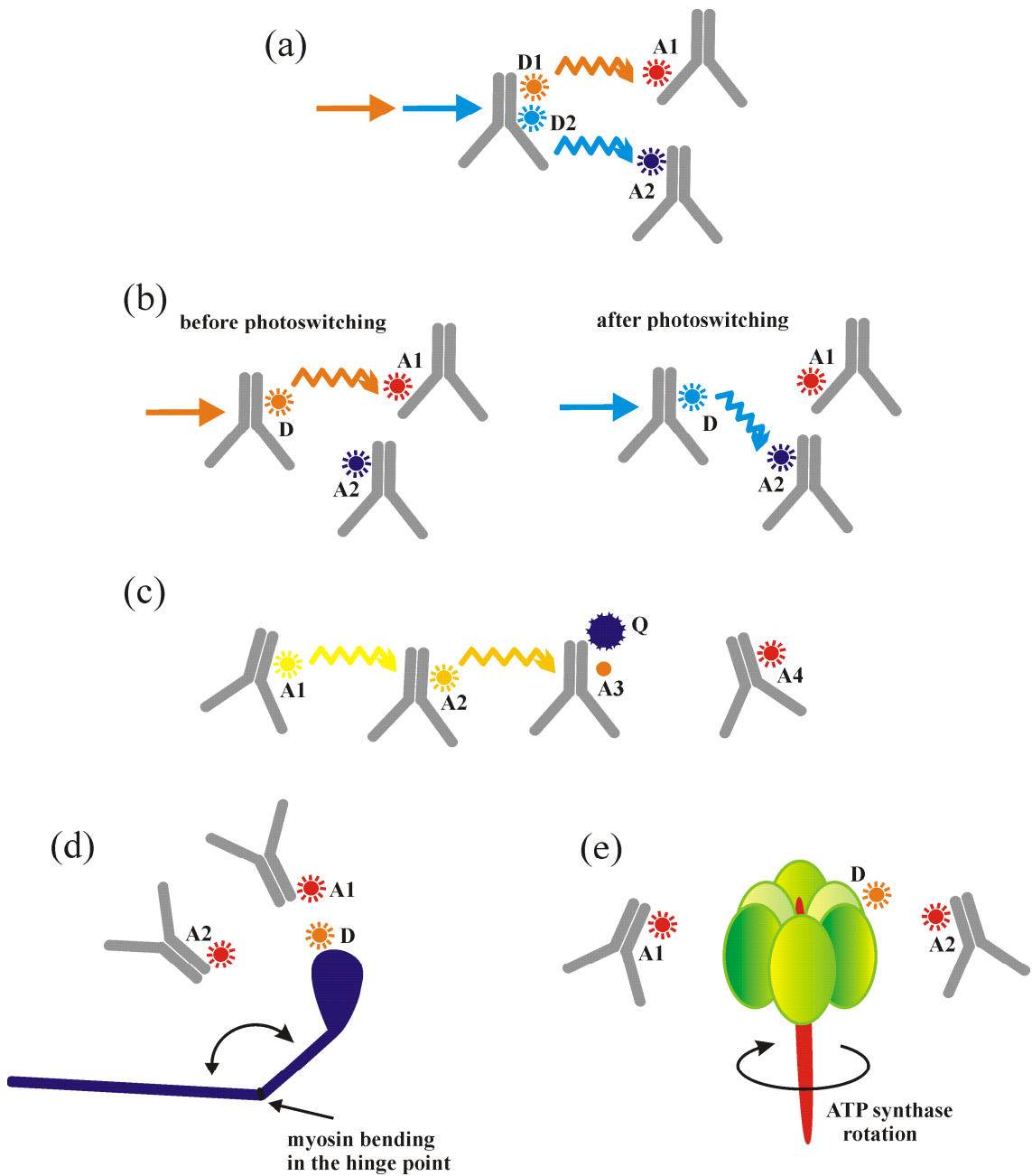

Figure 2. Routing mechanisms for FRET-based nanonetworks:

a) Transmitting nanomachine has 2 fluorophores working as nanoantennas. As the fluorophores have different absorption and emission spectra, signals can be transmitted via D1 to A1 or via D2 to A2.

b) Fluorophore D may be photoswitched and then its absorption and emission spectra change. Before photoswitching, it sends signals to A1; after photoswitching, it communicates with A2.

c) FRET signal may be passed in multiple hops from A1 to A3, but an active quencher Q blocks its further propagation. After deactivation of the quencher, the signal might also reach A4.

d) Myosin bends and the fluorophore D moves away from A1 and closer to A2. The transmission D-A1 is interrupted and the transmission D-A2 may be initiated.

e) ATP synthase rotates along its own axis. The fluorophore D attached to the ATP synthase may periodically communicate with A1 and A2.

protein (nanomachine) in various ways, usually utilizing active chemical groups present in the dye or in the protein. We can regulate not only the type of attached particles, but also their number, limited by the protein dimension. When the attached molecules have different absorption and emission spectra, they can receive and transmit signals of different frequencies, thus performing as nanorouters. This scenario resembles a situation where a wireless device is equipped with two or more antennas working in different frequency bands. Such a nanorouter is different from other solutions presented below in the sense that it does not need to be switched: it passes upcoming signals depending on their wavelength (Fig. 2a). The known fluorescent dyes that can be used to construct such a nanorouter are fluorophores from Alexa Fluor (AF), DyLight, and Atto families.

*Photoswitchable fluorophores*

Photoswitchable fluorophores, also called photoswitches, have been known in life sciences for years, but have recently gained significant attention as they allowed to develop various super-resolution imaging techniques [9]. What is important for nanorouting purposes, these fluorophores can be reversibly switched between two states by external electromagnetic impulses of a specific power and frequency. In each state, a reversible photoswitch has different absorption and emission characteristics, i.e. it is able to receive and transmit signals, also via FRET, in different frequency bands. A typical shift of the fluorophore emission or absorption spectrum is about 100 nm [10]. Therefore, photoswitchable fluorophores can be thought of as tunable antennas that, depending of an external control, are able to get FRET signals from and forward them to the chosen neighboring nanonodes. Certain connections in nanonetworks can be open and shut in this way (see Fig. 2b). The known reversible photoswitchable fluorophores are, among others, Cy5 molecules, numerous AF and Atto dyes.

*Quenchers*

A quencher works when being in close proximity to a fluorescent molecule called reporter; the quencher suppresses fluorescence of the reporter and in turn it blocks all its signals. The reporter can resume its transmissions when the quencher is removed or deactivated. This latter effect is usually realized via changes in the environment of the fluorophore (temperature, pH). For the communication applications, quenchers may be used in order to temporarily block FRET signals propagating in undesired directions, thus blocking the chosen links between nanonodes (see Fig. 2c). Multiple quenchers are commercially available, e.g. TAMRA or Dabcyl, but the most suitable are so-called dark quenchers, e.g. Black Hole Quenchers, which emit neither photons nor FRET signals.

*Proteins with changeable shape*

Interaction of two or more proteins with each other or a protein with a small molecule (e.g. an ion) may lead to the binding of these entities. This binding sometimes causes a conformational change of the protein, i.e., the change of its shape. Some of the conformations are especially attractive for the nanorouting purposes, as they change the separations between the fluorophores attached to the protein and other fluorophores located on neighboring nanomachines. The FRET efficiency decreases with the sixth power of the separation between the fluorophores, so via control of this separation one may in fact open or shut down chosen nanolinks. A good example is myosin which is a protein that, changes its shape, i.e., bends its part, after binding a $Ca^{2+}$ ion binding,. As a result, fluorophores attached to myosin may be put sufficiently close to other fluorophores attached to another nanomachine, just in order to perform a successful FRET transmission (see Fig. 2d). In this example, the $Ca^{2+}$ ion may be imagined as a trigger that activates the switch enabling a specific nanolink. It should be emphasized that the conformational change of myosin shape is reversible after the $Ca^{2+}$ ions are removed.

*ATP synthase*

Finally, a very promising protein is an ATP synthase (a motor enzyme creating ATP) which is able to rotate around its own axis (see Fig. 2e). This effect can be used to periodically communicate with some other nanomachines located nearby, either broadcasting the same signal to all the neighbors or dividing the signal among them in the time division multiple access (TDMA) manner. The rotation of the ATP synthase may be initiated providing electrical voltage or adding ATP.

To sum up, all these routing mechanisms, based on the properties of specific molecules, enable to control signals in nanonetworks in numerous different manners, mimicking the solutions well known from telecommunication networks. The signals can be routed to the chosen receivers, broadcasted or de-multiplexed, certain links may be switched on and off. All these techniques may be used in the same nanonetwork, offering a great variety of solutions, depending on the particular need.

EXPERIMENTAL STUDIES ON NANOROUTING

In order to present not only theoretical analysis, but also practical experience on nanorouting, we validated experimentally the first of the proposed routing mechanisms (proteins with a few different fluorescent dyes). Communication over nanometer distances occurring between devices of nanometer dimensions cannot be observed in classical telecommunication laboratories. Therefore, we performed experiments in a biophysical lab using a confocal microscope equipped with a FLIM (Fluorescence Lifetime Imaging Microscopy) module, which enabled to measure efficiency of the FRET process between nanonodes. First, we constructed a network composed of 3 nanonodes, as in Fig. 2a. The nodes were proteins: Immunoglobulin G with fluorophores attached working as FRET nanoantennas. In particular, we had one nanonode (rabbit anti-mouse IgG 610-

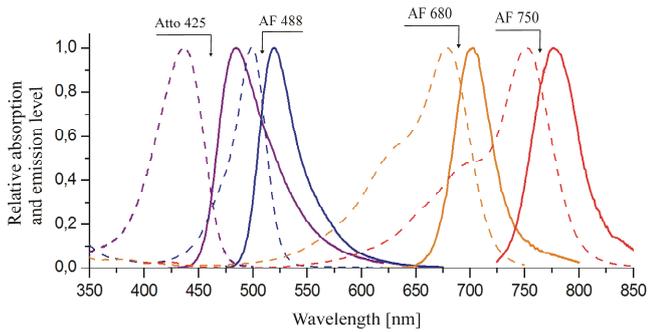

Figure 3. Absorption (dashed lines) and emission (solid lines) spectra of the fluorophores used in the experiment [11].

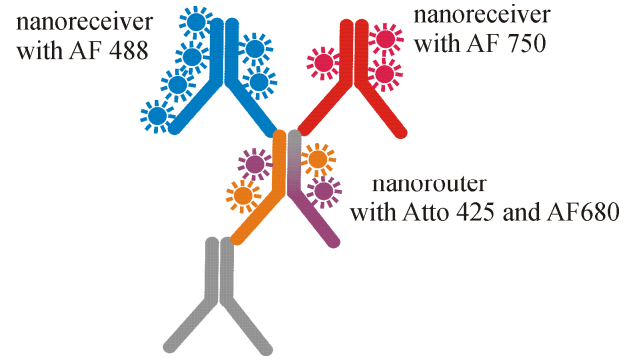

Figure 4. The experimental scenario: a chain of antibodies where one of them acts as nanorouter with Atto 425 and AF 680 dyes (violet and orange ones) and two others are nano receivers with AF 488 (blue) fluorophores and AF 750 (red) ones.

451-C46 produced by Rockland) operating as a nanotransmitter/nanorouter with two types of fluorophores attached to it: Atto 425[1] and AF 680. These two types of fluorophores are characterized by significantly different emission spectra and thus, if excited, may emit FRET signals which can be reached by different receivers. Apart from the nanorouter, we had two other nodes working as nanoreceivers: the one with AF 488 (goat anti-rabbit IgG ab150077 produced by Abcam) and the second one with AF 750 (goat anti-rabbit IgG A-21039 produced by Sigma). In this nanonetwork the emission spectra of Atto 425 matches the absorption spectra of AF 488 and similarly, the emission spectra of AF 680 matches the absorption spectra of AF 750, see Fig. 3. It means that we had two nanolinks created: the first one between Atto 425 and AF 488 (from the nanorouter to the first nanoreceiver), and the second one between AF680 and AF750 (from the nanorouter to the second nanoreceiver).

As we could not exactly control the movements of the proteins in the laboratory samples, the nanonodes (Immunoglobulin G molecules) were fixed to a large chain of DNA sequence and other proteins, in order to keep the transmitting and receiving nanoantennas (fluorophores) about 8–15 nm from each other, see Fig. 4. The fixing was done only for the measurement purposes; in a real scenario the nanonodes could float freely in their environment (e.g. in a human cell).

The laboratory experiments were performed on fixed HeLa 21-4 cells (a common cell line used in laboratory research). In the nucleus of each cell there were about $3 \cdot 10^6$ nanonetworks (the nanorouter and both nanoreceivers) attached to the DNA sequence. We analyzed the data from 57 nuclei, so we had over $10^8$ nanonetworks examined, which was a statistically credible number. The cells containing the nanonetworks were exposed to short pulses of laser light. The wavelength of light utilized to excite fluorophores attached to the nanotransmitter, Atto 425 and AF 680, were 405 and 640 nm, respectively. During each laser pulse, only few fluorophores (donors) were excited in the whole sample. It means that, despite having multiple fluorophores at the nanorouter and at the nanoreceivers (Fig. 4), we investigated a scenario of a single donor and multiple acceptors which was like a SIMO (single-input multi-output) channel. In real applications, all the donor molecules may be excited at once using full MIMO communication, which can additionally increase the FRET efficiency comparing with the values shown in this article [7–8].

The main purpose of the experiment was to measure how much of the FRET signal could be passed from the nanorouter to the chosen nanoreceiver. When a donor in an excited state is close to an acceptor, and the emission and absorption spectra of these molecules are matched, energy may be passed from the donor to the acceptor via FRET. In a laboratory sample, FRET can be observed indirectly as a decrease of donor fluorescence intensity (or an increase of acceptor fluorescence intensity). Thus, measurements of fluorescence intensity can give information regarding the efficiency of the FRET transfer. An alternative way to obtain the value of FRET efficiency is to monitor the fluorescence lifetime of the donor. Fluorescence lifetime is defined as the mean time the donor spends in its excited state, before returning to the ground state. Since FRET represents an additional way for the donor to depopulate its excited state, the lifetime of the donor decreases in the presence of an acceptor. Comparing the lifetime of the donor (nanorouter) in the absence and presence of the acceptor (nanoreceiver) allows to calculate the FRET efficiency [12]. As the latter approach is less prone to photobleaching effects than intensity-based methods, we have chosen it for our experiments.

The measurements were conducted using a Leica TCS SP5 II SMD confocal microscope (Leica Microsystems GmbH) integrated with FCS/FLIM module from PicoQuant GmbH. The FLIM module enabled to combine fluorescence lifetime measurements with optical microscopy. The lifetime measurements were performed in Time Correlated Single Photon Counting mode. The nanonetworks located in the nuclei of HeLa cells were exposed to a pulse laser exciting in turns the molecules of Atto 425 and AF 680. The detection of photons emitted by these molecules was observed with Single

---

[1] The numbers of fluorophores, 425, 488, 680 and 750, indicate the wavelengths of their maximum absorption, given in nanometers. Their emission spectra are usually shifted by few dozen of nanometers into higher wavelengths.

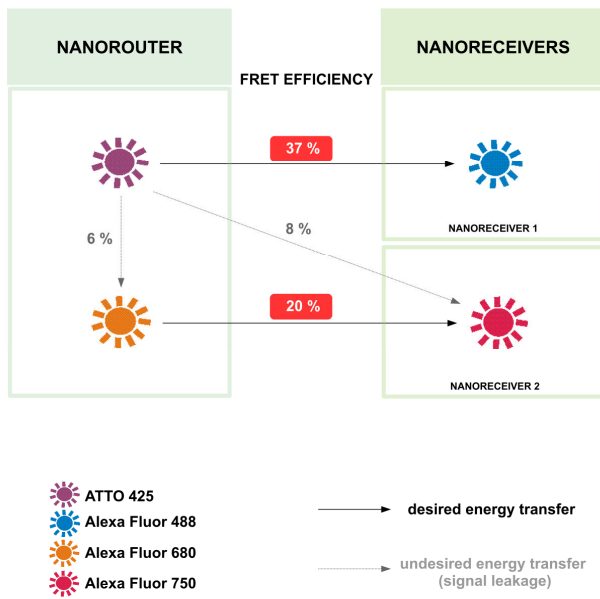

Figure 5. The FRET efficiency values characterizing the signal transmission between the nanorouter and the nanoreceivers.

Photon Avalanche Diode detectors. The laser repetition rate was 40 MHz while the laser power was adjusted to obtain a photon counting rate of 200–300 kCounts/s. The acquisition time for each field of view (image) was set to 1 minute. The analysis of the results was done with SymPhoTime II software using tail fitting of the lifetime functions. Lifetime decays of Atto 425 and AF 680 were fitted with a two-exponential function, i.e. a sum of two exponential functions with different amplitudes and lifetimes. For the calculation of FRET efficiency, the average lifetime was used. The goodness-of-fit was estimated based on the weighted residuals and the chi squared value.

The results of performed experiments are summarized in Fig. 5. The FRET values in the pairs Atto 425 → AF 488 and AF 680 → AF 750 show how efficient the transmission between the nanorouter and the nanoreceivers might be. During the measurements, due to the limited sensitivity of the detectors, it was not possible to excite multiple nanorouter donors at once. Thus, in practice, we have measured the FRET efficiencies for the single-input multiple-output (SIMO) case, which is much less effective than the full MIMO-FRET. This is why the FRET values are rather low for the telecommunication purposes. Fortunately, using MIMO-FRET and suitable coding techniques with enough redundancy, the FRET-based nanocommunications can guarantee bit error rate at the level of $10^{-6}$ [8]. The results also raise an issue of undesired signal leakages: when the donor Atto 425 at the nanorouter is excited, 37 percent of the signal is transferred to the AF 488 at the first nanoreceiver, but at the same time 8 percent passes to AF 750 at the second nanoreceiver. The same problem occurred during additional control measurements where we observed another signal leakage between the donors at the nanorouter (Atto 425 and AF 680) at the level of 6 percent. Both leakages are caused by the partial and undesired overlap of the Atto 425 emission spectrum and AF 680/750 absorption spectra, see Fig. 3.

## OPEN ISSUES

While the article proposes five new routing techniques for nanonetworks and reports the experimental validation for one of them, it is clear that this research area is at the beginning of its road and many questions remain unanswered. Below we present the most important open issues.

### Routed signal leakages

As reported in the previous section, the separation between the links outgoing from the nanorouter is far from being perfect: it may happen that the signal from the nanorouter is absorbed at the wrong nanoreceiver. The reason is that the receiving molecules are characterized by broad absorption spectra and may absorb signals emitted at wavelengths even 300 nm shorter than their maximum absorption wavelength (see again Fig. 3). The most straightforward solution of this problem is to increase the wavelength separation between the pairs of fluorophores used for communication purposes. Because of the fluorescence spectroscopy requirements, the fluorophores currently available at the market are working in the wavelength range of 300–900 nm, which is in general the visible and partially infrared light range. The molecules with emission/absorption spectra at higher and lower wavelengths would, at least partially, solve this problem. Another possibility would be to use fluorophores with very narrow emission spectra, e.g. BODIPY dyes.

### Switching time

While the propagation speed for FRET signals is quite high, the routing itself may not be so fast. The routing mechanisms proposed in this article require, in some cases, change in the shape of a molecule (e.g. myosin) or in its properties (photoswitches, quenchers). The time of this change, i.e. the nanorouter switching time, may even reach several dozens of seconds (e.g. in the case of fluorophores photoswitching) [13], which is too long for most of telecommunication purposes. There is a clear need for shortening the switching time. It may be done by finding more sensitive molecules that could act as nanorouters or carefully choosing the switching conditions (pH of the environment, radiation intensity, and frequency, etc).

### Transmitter excitation

In the current FRET experiments, the transmitters (donors) are excited by an external laser impulse. While it may be the case of some nanonetworking applications, in most of them it should be the nanomachine itself that initiates the communication. In such a situation, BRET (Bioluminescence Resonance Energy Transfer) may be applied instead of FRET.

In BRET, the excitation energy comes not from an external source, but from a chemical reaction taking place near the donors. Thus, when a nanomachine is going to send a signal, it induces the chemical reaction which, in turn, excites the donors. The manner how the nanomachine initiates this reaction still remains an open issue.

*Signal storage*

The FRET phenomenon enables communication between the fluorophores, but these molecules cannot hold the received signal. Before releasing its energy (emitting a photon, via FRET or other processes) and returning to the ground state, a molecule spends no longer than a few dozens of nanoseconds in the excited state. This creates a problem for the considered nanonetworks, as the nanonodes should have some buffers where the signals could be stored, until the nodes decide to resend them further. Currently, there is no clear solution for creating these buffers. One option could be using phosphorescent molecules that are able to store energy even 9 orders of magnitude longer than fluorophores [14]. Phosphorescence is a phenomenon similar to fluorescence, but its excitation state lifetime can reach minutes, as releasing energy by a phosphorescent molecule is related to a transition, which is, according to the quantum mechanics, forbidden. Storing signals in phosphorescent molecules could be even more important when using nanorouters with long switching time, as mentioned two subsections earlier.

*Nanomachine movements*

There is also a general issue of nanomachine movement control. While it is a little out of scope of this article, it should be noted that relative nanomachine positions and their separation are crucial for the efficiency of FRET-based communication, as it depends on sixth power of the transmitter-receiver (donor-acceptor) distance. Thus, manipulating the nanomachines, e.g. putting them into a closer range for the time of communication will result in much higher efficiency of information transfer. It can be done using proteins with changeable shapes, as indicated in the section about routing mechanisms, but there are some other options possible, like bridging of free monomers of IgE or JAK receptors [15].

## CONCLUSION

The phenomenon of FRET seems to be a very promising solution for nanocommunications, with a very high propagation speed, comparing with other techniques proposed so far. While point-to-point transmission with FRET has been already investigated, both theoretically and experimentally, this article expands the topic analyzing possibilities of routing FRET signals through nanonetworks. The new proposed routing techniques rely on physical and biological properties of specific molecule types, mainly proteins and fluorophores. With current development of biotechnology, these molecules may be easily manipulated and put together in large, static or mobile, structures. The routing techniques enable maintaining end-to-end communication in whole nanonetworks, which is a crucial step to have future tiny nanodevices working together on complex endeavors.

## ACKNOWLEDGMENT

Authors would like to thank P. Cholda, A. Jajszczyk, A. Lason, and R. Wojcik for reviewing the manuscript and their constructive comments. The work was performed under the contract 11.11.230.018 and also funded by the National Science Centre based on the decision number DEC-2013/11/N/NZ6/02003. The confocal microscope was purchased through an EU structural funds grant BMZ no. POIG.02.01.00-12-064/08.

PAWEL KULAKOWSKI (kulakowski@kt.agh.edu.pl) received a Ph.D. in telecommunications from the AGH University of Science and Technology in Krakow, Poland, in 2007. Currently he is working there as an Assistant Professor. He was also working few years in Spain, as a visiting post-doc or professor at Technical University of Cartagena, University of Girona, University of Castilla-La Mancha and University of Seville. He co-authored about 30 scientific papers, in journals, conferences and as technical reports. He was involved in numerous research projects, especially European COST Actions: COST2100, IC1004 and CA15104 IRACON, focusing on topics of wireless sensor networks, indoor localization and wireless communications in general. His current research interests include molecular communications and nanonetworks. He was recognized with several scientific distinctions, including 3 awards for his conference papers and a scholarship for young outstanding researchers.

KAMIL SOLARCZYK (kj.solarczyk@uj.edu.pl) received M.Sc. and Ph.D. degrees in biophysics from the Jagiellonian University, Krakow, Poland, in 2010 and 2016, respectively. He is a postdoctoral researcher at Department of Cell Biophysics, Faculty of Biochemistry, Biophysics and Biotechnology, Jagiellonian University. His research interests include the DNA repair processes, chromatin architecture and nanoscale communications.

KRZYSZTOF WOJCIK (krzysztof.wojcik@uj.edu.pl) received his M.Sc. and Ph.D. degrees in biophysics from the Jagiellonian University in Krakow, Poland 2003 and 2015, respectively, and an M.D. from the Jagiellonian University Medical College in Krakow, Poland 2007. He was an Assistant at Division of Cell Biophysics Faculty of Biochemistry, Biophysics and Biotechnology Jagiellonian University (2007-2014). He is an Assistant at Allergy and Immunology Clinic in II Chair of Internal Medicine JUMC. His research interests include confocal microscopy techniques and their applications in autoantibodies research, as well as the use of fluorescent probes in nanocommunications.